\documentclass[%
reprint,superscriptaddress,amsmath,amssymb, floatfix,aps,prl,twocolumn,showpacs]{revtex4-2}
\usepackage{amsmath,amssymb,color,bm,graphicx}
\usepackage[colorlinks=true,linkcolor=blue,citecolor=blue,urlcolor=red]{hyperref}
\usepackage{caption}
\usepackage{subcaption}
\usepackage{orcidlink}
\usepackage[normalem]{ulem}  

\begin{document}

\title{Suprathermal bulk viscosity in binary neutron star  mergers}

\author{Pranjal Tambe~\orcidlink{0000-0003-2293-6953}}
~\email{pranjal.tambe@iucaa.in}
 \affiliation{%
 Inter University Centre for Astronomy and Astrophysics, Ganeshkind, Pune 411007, India
}%
\author{Debarati Chatterjee~\orcidlink{0000-0002-0995-2329}}%
~\email{debarati@iucaa.in}
\affiliation{%
 Inter University Centre for Astronomy and Astrophysics, Ganeshkind, Pune 411007, India
}%
\begin{abstract}
Numerical relativity simulations of binary neutron star mergers and core-collapse supernovae indicate that the neutron star matter can be driven far from equilibrium due to large amplitude density oscillations, with frequency in the kHz range. Urca processes in neutron star matter try to restore the system towards equilibrium providing strong bulk viscous damping of these density oscillations. Standard bulk viscosity calculations typically assume small amplitude density oscillations, such that the deviation from chemical equilibrium $\mu_{\Delta}$ is less than the temperature $T$. In this work we compute the complete amplitude dependent bulk viscosity, taking into account, for the first time, thermal blurring effects and collisional broadening of in-medium nucleons. Our results show that the non-linear effects in bulk viscosity start to become important when the oscillation amplitude is about $10$ percent of the background density. In this regime the bulk viscosity can increase by three to five orders of magnitude. These results have important consequences for simulations of binary neutron star mergers as well as their multi-messenger observations.
\end{abstract}
\maketitle

\section{Introduction}
\label{sec:intro}
Both isolated Neutron Stars (NS) and binary NS mergers can be important sources of gravitational waves. The violent dynamics in binary neutron star (BNS) mergers and core-collapse supernovae can drive the neutron star matter out of flavor equilibrium because of large-amplitude density oscillations in the kHz frequency range \cite{Janka:2006fh,Lentz:2015nxa,Baiotti:2016qnr, Alford:2017rxf,Hanauske:2019qgs, Burrows:2020qrp, Hammond:2021vtv, Most:2022wgo}. In nuclear matter composed of neutrons, protons and electrons (npe), the flavor re-equilibration is established via the weakly interacting Urca processes. This re-equlibration can produce strong bulk viscous damping of the density oscillations \cite{Alford:2017rxf, Alford:2019qtm, Alford:2020lla, Most:2021zvc, Alford:2021lpp, Alford:2022ufz, Most:2022yhe, Alford:2023uih}. It has been seen in numerical simulations that large bulk viscosities effectively damp the violent density oscillations within stellar cores right after the merger, shifting the peak frequency of the post-merger gravitational wave spectrum by up to $100$~Hz \cite{Chabanov:2023blf}. It has also been shown that the inclusion of large bulk viscosity in BNS merger simulations can heavily suppress dynamical mass ejection, reducing the total ejecta mass by up to a factor of five \cite{Chabanov:2023abq}. This can have significant implications for multi-messenger observations of the electromagnetic counterparts to BNS merger events detected in gravitational waves~\cite{Radice:2018xqa, Zappa:2022rpd, Margalit:2022rde, Chabanov:2023abq}.

Numerical relativity simulations of BNS mergers indicate that within the first few milliseconds of the merger, density oscillations can have large amplitude \cite{Alford:2017rxf, Most:2022yhe}. These large-amplitude oscillations can push the system far from its equilibrium configuration into the suprathermal regime, where the deviation from chemical equilibrium ($\mu_{\Delta}$) can be larger than the temperature ($T$), i.e., $|\mu_{\Delta}|/T \geq1$. The relevance of the suprathermal regime was highlighted in \cite{Hammond:2021vtv}, where it has been shown that $|\mu_{\Delta}|/T \geq1$ in large parts of the postmerger remnant, especially in the inner core of the remnant. These findings suggest that it is crucial to incorporate non-linear effects in bulk viscosity to consistently compute the amplitude dependent viscous damping of these density oscillations \cite{Alford:2010gw}.
For neutron stars composed of nucleonic matter, chemical equilibration is primarily driven by Urca processes \cite{Haensel:2000vz, Haensel:2001mw}. The standard bulk viscosity calculations rely on computing the equilibration timescales from two distinct Urca channels. Due to momentum conservation, the direct Urca (dUrca) process is the dominant equilibration mechanism only above the direct Urca threshold density, where the proton fraction $x_p\geq0.11$, while it remains kinematically suppressed below this density. The modified Urca (mUrca) process dominates equilibration below the direct Urca threshold density. A recent reformulation for calculating the Urca reaction rates incorporating thermal blurring of Fermi surfaces and in-medium collisional broadening of nucleons was proposed in~\cite{Alford:2024xfb}, which allow for computing the total reaction rates across the threshold density within a single unified framework called the Nucleon Width Approximation (NWA). This framework was applied recently by the authors of this work to compute amplitude-independent bulk viscosity in the subthermal regime, and for the first time, also incorporating the effects of a strong magnetic field, assuming the oscillation amplitudes are sufficiently small that the deviation from chemical equilibrium satisfies $\mu_{\Delta}\ll T$ \cite{Tambe:2025evw}.
 \\
In this work, we develop a novel methodology by extending our analysis to the non-linear suprathermal regime, in order to compute the amplitude-dependent bulk viscous dissipation of large-amplitude density oscillations, incorporating thermal effects and collisional broadening of in-medium nucleons using the NWA framework. The results thus obtained are highly relevant for improving numerical simulations of BNS mergers and core-collapse supernovae and also have important implications for gravitational wave detections and their electromagnetic follow-up observations.
\\
The paper is structured as follows. In Sec.~\ref{sec:formalism}, we introduce the formalism used in this investigation. For this study, we consider the neutron star core to be composed of neutrons, protons, and electrons. We assume that neutrinos are not trapped within the fluid element. In Sec.~\ref{sec:results} we present the results of our calculations and we discuss their broad implications in Sec.~\ref{sec:discussions}.

\section{Formalism}
\label{sec:formalism}

In order to compute the bulk viscous dissipation from large amplitude oscillations, we follow the formalism introduced in~\cite{Alford:2010gw}, which is briefly recapitulated below. The density oscillation around the equilibrium value $\bar{n}_*$ is:
\begin{equation}
    n_*=\bar{n}_*+\Delta n_*\sin(\omega t),
\end{equation}
where $\omega$ is the angular frequency of the oscillations. The density oscillations induce chemical potential fluctuations given by:
\begin{equation}
    \delta\mu_{\Delta}=\left.\frac{\partial\mu_{\Delta}}{\partial n_*}\right|_x\delta n_* + \left.\frac{\partial\mu_{\Delta}}{\partial x}\right|_{n_*}\delta x ,
    \label{eq:diff_mu}
\end{equation}
 where $\mu_{\Delta}=\mu_n-\mu_p-\mu_e-\mu_I\gg T$ is  the chemical potential imbalance, $\mu_i$ is the chemical potential for particle $i=n,p,e$ and $\mu_I$ is the correction to the cold beta equilibrium condition at finite temperature. Here $x$ is the particle fraction that changes during the oscillation cycle, i.e., the proton fraction. This gives the expression,
 \begin{equation}
     \frac{d\mu_{\Delta}}{dt} = C\omega\frac{\Delta \bar{n}_*}{\bar{n}_*} \cos(\omega t)+B \bar{n}_* \frac{dx}{dt},
 \end{equation}
 where $B,C$ are the susceptibilities calculated from the Equation of State (EoS) defined as:
\begin{equation}
    C=\bar{n}_* \left.\frac{\partial\mu_{\Delta}}{\partial n_*}\right|_x , \quad B=\frac{1}{\bar{n}_*} \left.\frac{\partial\mu_{\Delta}}{\partial x}\right|_{n_*} .
    \label{eq:suscept}
\end{equation}
The net equilibration rate is given by,
\begin{equation}
    \Gamma^{(\leftrightarrow)} = \Gamma^{(\rightarrow)}-\Gamma^{(\leftarrow)}=\bar{n}_*
\frac{\partial x}{\partial t}~.
\end{equation}
To calculate the amplitude dependent bulk viscosity, we need to expand the net reaction rate in powers of fluctuations from chemical equilibrium $\mu_{\Delta}/T$ in the form:
\begin{equation}
    \Gamma^{(\leftrightarrow)} = -\widetilde{\Gamma}T^{\kappa}\mu_{\Delta}\left(1+\sum_{j=1}^N\chi_j \left(\frac{\mu_{\Delta}}{T}\right)^j\right)\,.
    \label{eq:rates_expansion}
\end{equation}
In contrast to \cite{Alford:2010gw}, we expand the net reaction rate in all powers of $\mu_{\Delta}/T$, instead of just the odd powers, as the thermal blurring and collisional broadening effects modify the phase space available for the reactions.
Then, in terms of dimensionless variables 
\begin{equation}
    \varphi \equiv \omega t,\hspace{0.5cm}A(\varphi)\equiv\frac{\mu_{\Delta}}{T},
\end{equation}
we can rewrite the differential equation for chemical potential fluctuations Eq.~\ref{eq:diff_mu} as:
\begin{equation}
    \frac{dA}{d\varphi}=d\,cos(\varphi) -fA\left(1+\sum_{j=1}^N\chi_j A^{2j}\right)\,,
    \label{eq:diff_A}
\end{equation}
where
\begin{equation}
    d=\frac{C}{T}\frac{\Delta n_{*}}{n_{*}},\hspace{0.5cm}f=\frac{B\widetilde{\Gamma}T^{\kappa}}{\omega}.
    \label{eq:prefactors}
\end{equation}
Here $d$ is the driving term which depends on the amplitude of induced oscillations, and $f$ is the feedback term that depends on the rates of equilibrating processes and $B,C$ as defined in Eq.~\ref{eq:suscept}.
The amplitude dependent bulk viscosity is then given by Eq.~($22$) from \cite{Alford:2010gw}:
\begin{equation}
    \zeta=\frac{TC}{\pi\omega B}\frac{\bar{n}_*}{\Delta {n}_*} \int_0^{2\pi}A(\varphi;d,f) \cos(\varphi)d\varphi,
    \label{eq:BV_general}
\end{equation}
where $A(\varphi;d,f)$ is the solution of Eq.~(\ref{eq:diff_A}). This can also be written as
\begin{equation}
    \zeta = \frac{C^2}{2\omega B}\mathcal{I}(d,f),
\end{equation}
where 
\begin{equation}
    \mathcal{I}(d,f)= \frac{2}{\pi d}\int_0^{2\pi}A(\varphi;d,f) \cos(\varphi)d\varphi.
    \label{eq:integral}
\end{equation}

To compute suprathermal bulk viscosity, we need to expand rates in powers of $A=\frac{\mu_{\Delta}}{T}$, as in Eq.~(\ref{eq:rates_expansion}). We then have to solve for $A$ from Eq.~(\ref{eq:diff_A}) and this solution is then to be substituted into the general expression for bulk viscosity in Eq.~\ref{eq:BV_general} or~\ref{eq:integral}.




\section{Results}
\label{sec:results}

In this study, we will use two finite temperature equations of state (EoS): (i) IUF, which has a direct Urca threshold at $n_B=4 n_0$, where $n_0=0.16$ fm$^{-3}$ is nuclear saturation density, and (ii) QMC-RMF3, which has no direct Urca threshold within the density of our interest. 
In order to compute the amplitude-dependent bulk viscosity from Eq.~\ref{eq:BV_general}, we need to express our numerical calculations for rates in NWA formalism in the form shown in Eq.~\ref{eq:rates_expansion}. This implies we need to study the behavior of equilibration rates as a function of temperature around the true chemical equilibrium, where $\mu_{\Delta}=\mu_n-\mu_p-\mu_e-\mu_I=0$. We also need to express the rates as a polynomial expansion in the factor $\mu_{\Delta}/T$.

For this purpose, we first show a plot for the net reaction rates as a function of deviation from the true chemical potential $\mu_{\Delta}/T$ in Fig.~\ref{fig:polyfit_rates} in Appendix~\ref{sec:appendix_a}. We also show a polynomial fit to express the net reaction rates in the form of Eq.~\ref{eq:rates_expansion}. In \cite{Alford:2010gw}, the expression for net reaction rates was expanded in odd powers of $\mu_{\Delta}/T$. When we compute the net reaction rates incorporating thermal blurring and collisional broadening effects, we see from Fig.~\ref{fig:polyfit_rates} that an expansion of net reaction rates only in odd powers of $\mu_{\Delta}/T$ does not accurately capture the effects at large $\mu_{\Delta}/T$. We thus use a polynomial expansion in all powers of $\mu_{\Delta}/T$ for the net reaction rates to accurately capture the thermal blurring and collisional broadening effects.

\begin{figure}
    \centering
    \includegraphics[width=0.85\linewidth]{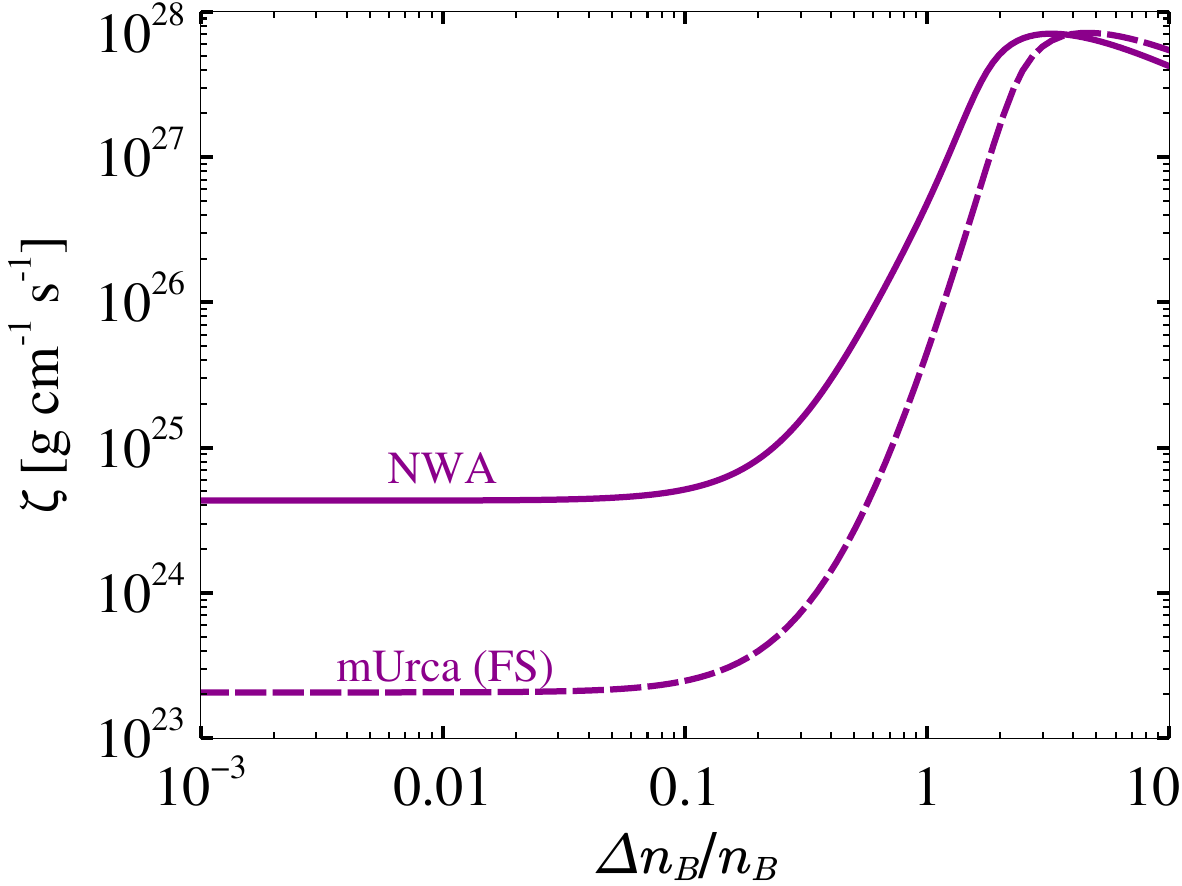}
    \caption{Comparison of bulk viscosity $\zeta$ in the suprathermal regime obtained from the reaction rates evaluated in the NWA formalism and the Fermi-surface (FS) approximation for the modified Urca process from \cite{Haensel:2001mw}, as a function of the amplitude of density oscillations for the IUF EoS at $n_B=n_0$.} 
    \label{fig:BV_supra_T1_nb1_IUF}
\end{figure}

\begin{figure}
    \centering
    \begin{subfigure}{0.45\textwidth}
    \includegraphics[width=\textwidth]{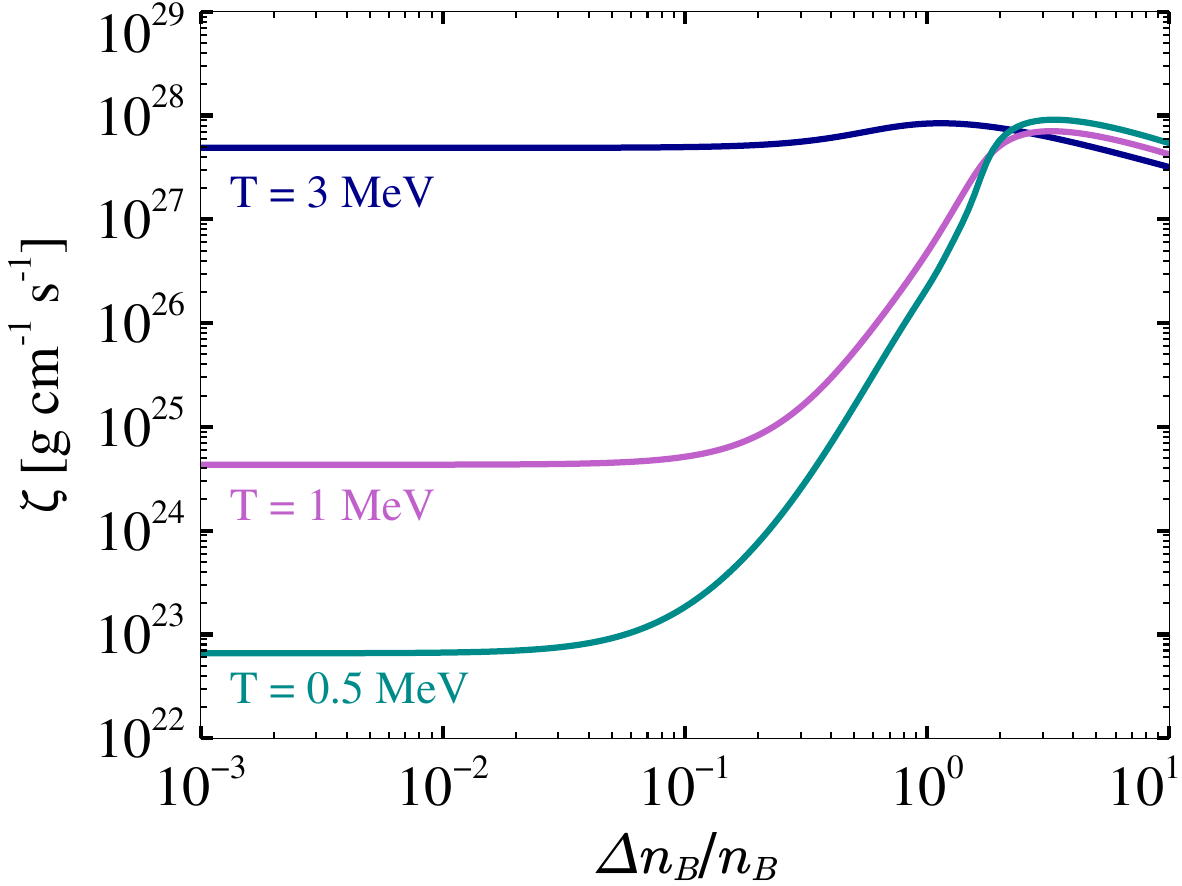}
    \caption{$n_B=n_0$}
    \label{fig:BV_supra_nb1_IUF}
     \end{subfigure}
     \\
    \begin{subfigure}{0.45\textwidth}
    \includegraphics[width=\textwidth]{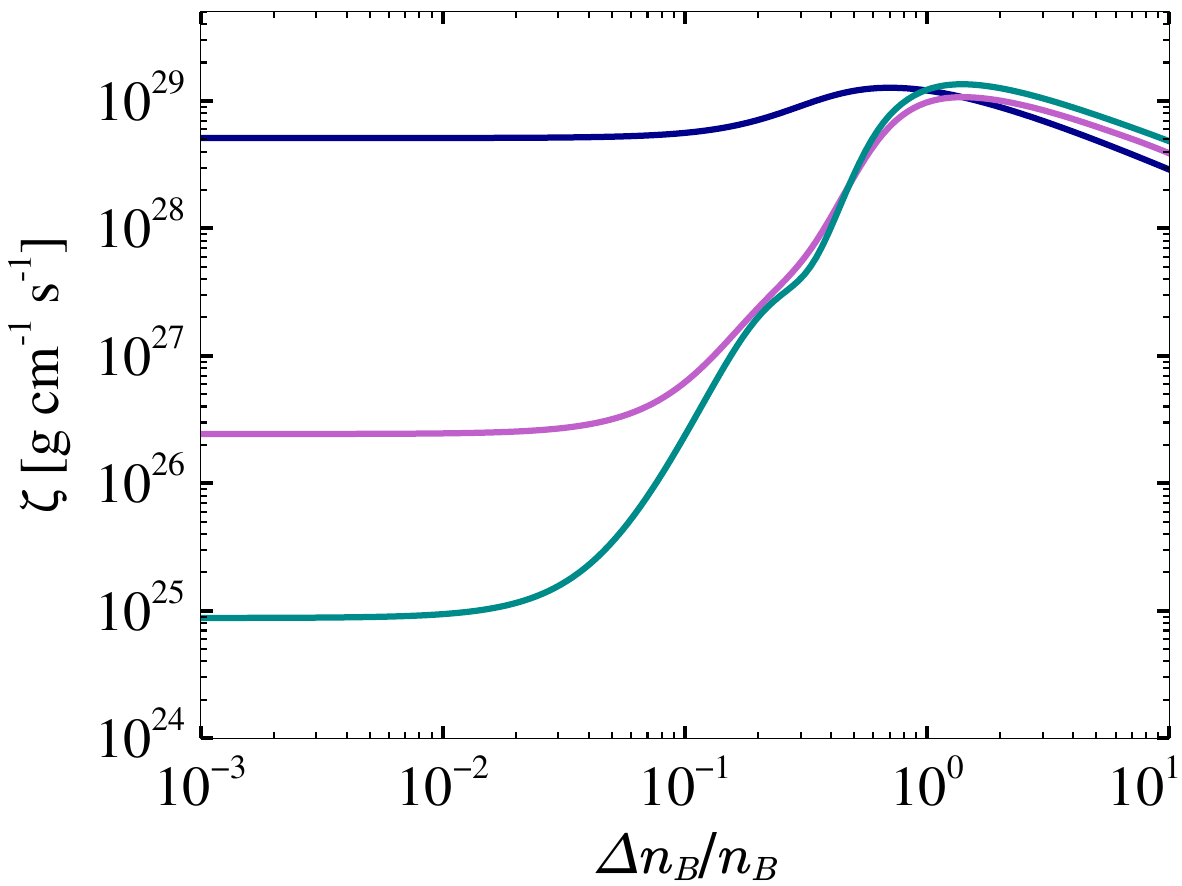}
    \caption{$n_B=4n_0$}
    \label{fig:BV_supra_nb4_IUF}
    \end{subfigure}
    \\
    \begin{subfigure}{0.45\textwidth}
    \includegraphics[width=\textwidth]{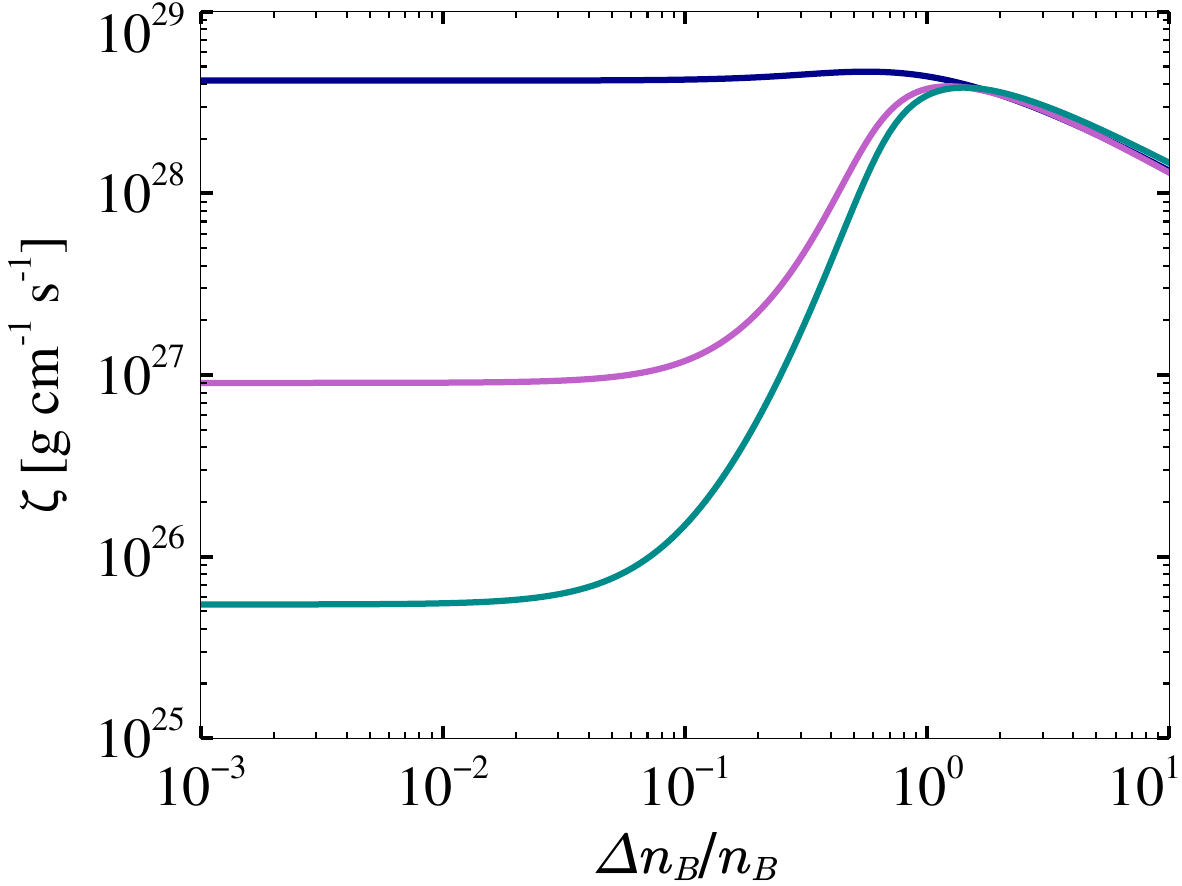}
    \caption{$n_B=6n_0$}
    \label{fig:BV_supra_nb6_IUF}
    \end{subfigure}
    \caption{
   Bulk viscosity $\zeta$ in the suprathermal regime, as a function of the amplitude of density oscillations for IUF EoS at $n_B=n_0,4n_0,6n_0$. The direct Urca threshold density for IUF EoS is $n_B=4.1n_0$.}
    \label{fig:tflavor}
\end{figure}
Using the polynomial fit for rates as a function of $\mu_{\Delta}/T$, we can calculate the bulk viscosity in the suprathermal regime using Eqs.~\ref{eq:diff_A} and \ref{eq:BV_general}. We first show a comparison between the bulk viscosity in the suprathermal regime, obtained by evaluating the reaction rates in the NWA-formalism used in this paper, with the calculations of modified Urca process in the Fermi-surface (FS) approximation \cite{Haensel:2001mw} in Fig.~\ref{fig:BV_supra_T1_nb1_IUF} for IUF EoS at $n_B=n_0$ and $T=1$ MeV. We see that the bulk viscosity in the subthermal limit is about an order of magnitude higher and it is larger even beyond amplitudes $\Delta n/n =1$. This can lead to an earlier damping of high amplitude density oscillations in neutron star mergers.

We now display the results for bulk viscous damping in the suprathermal regime as a function of amplitude of oscillation of frequency $f=1$ kHz, at different density and temperatures for IUF EoS in Figs.~\ref{fig:BV_supra_nb1_IUF}, \ref{fig:BV_supra_nb4_IUF} and \ref{fig:BV_supra_nb6_IUF}. We see that at different densities and at low temperatures, the non-linear effects in bulk viscosity tend to appear when the density amplitude $\Delta n/n$ is around 0.1, i.e. when the oscillation amplitude is around $10$ percent of the density of the fluid element. 

We see from Fig.~\ref{fig:BV_supra_nb1_IUF} that at density below the direct Urca threshold $n_B=n_0$ and low temperature $T=0.5$ MeV, the bulk viscosity increases with the amplitude of oscillation by up to five orders of magnitude. At $T=1$ MeV, the bulk viscosity increases with the oscillation amplitude by three orders of magnitude. As the temperature increases to $3$ MeV, the bulk viscosity stays close to its subthermal limit even for large oscillation amplitudes. From Fig.~\ref{fig:BV_supra_nb6_IUF} we see that above the direct Urca threshold density, where the direct Urca processes are the dominant equilibration channel, the bulk viscosity increases by up to three orders of magnitude at $T=0.5$ MeV and by an order of magnitude at $T=1$ MeV in the suprathermal regime.

\begin{figure}
    \centering
    \begin{subfigure}{0.45\textwidth}     \includegraphics[width=\textwidth]{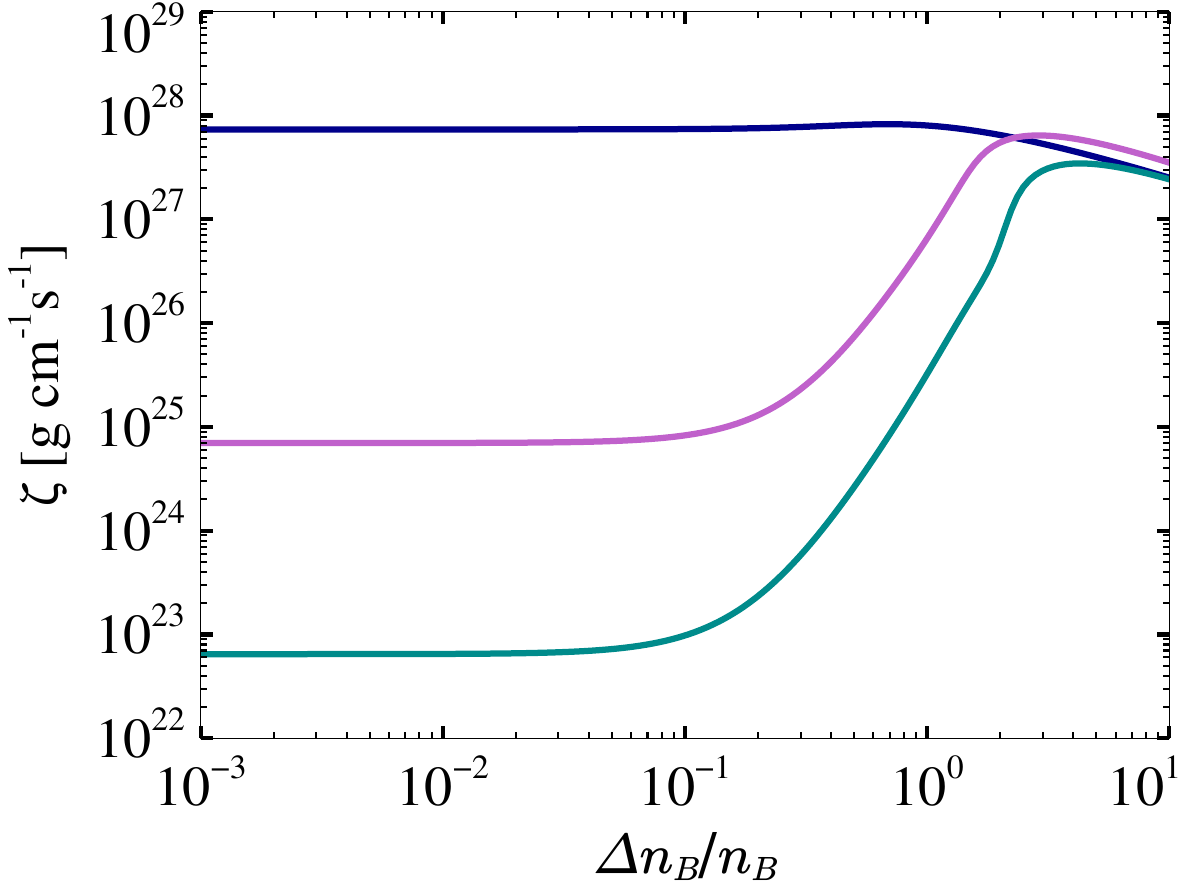}
    \caption{$n_B=n_0$}
    \label{fig:BV_supra_nb1_QMC}
    \end{subfigure}
    \\
    \begin{subfigure}{0.45\textwidth}     \includegraphics[width=\textwidth]{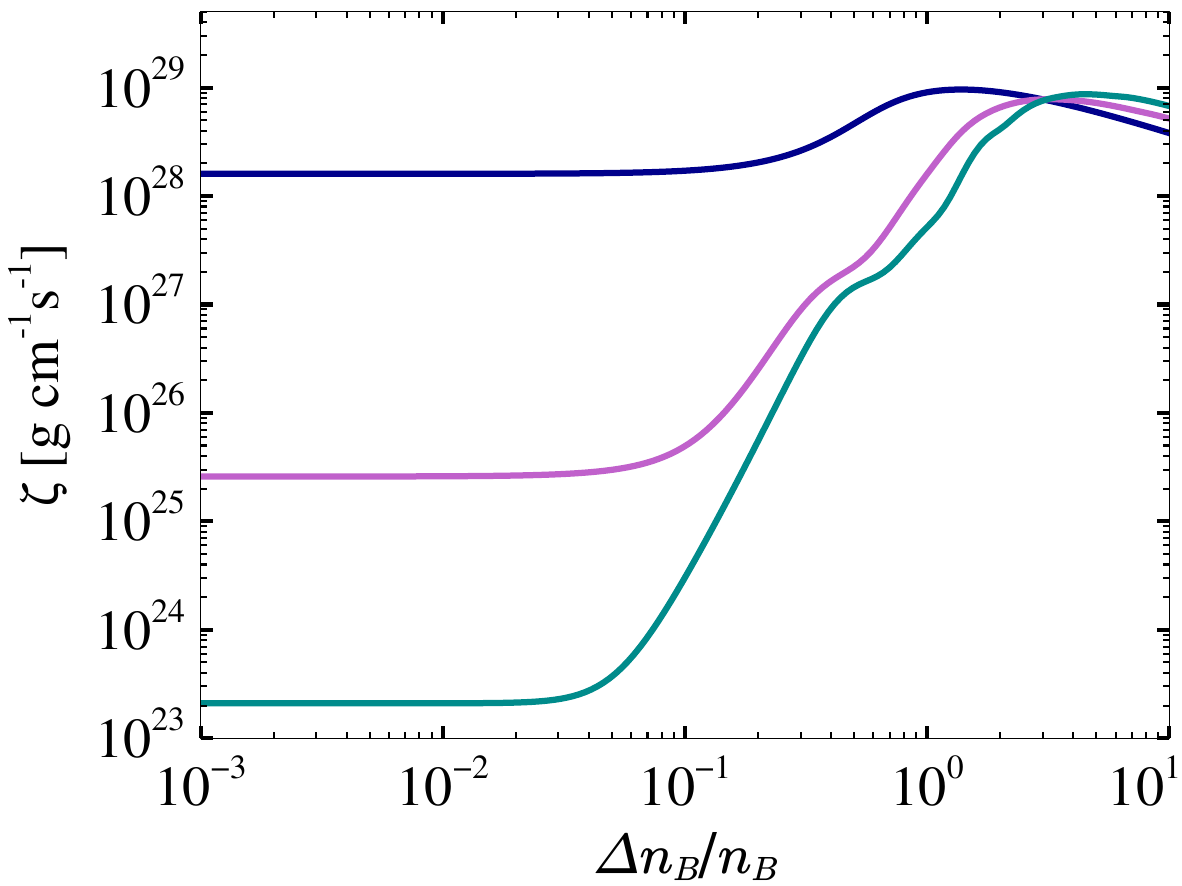}
    \caption{$n_B=6n_0$}
    \label{fig:BV_supra_nb6_QMC}
    \end{subfigure}
    \caption{Bulk viscosity $\zeta$ in the suprathermal regime, as a function of amplitude of density oscillations for QMC-RMF3 EoS at $n_B=n_0,6n_0$.}
\end{figure}

We now repeat the same calculations for QMC-RMF3 EoS, and the results are displayed in Figs.~\ref{fig:BV_supra_nb1_QMC}, \ref{fig:BV_supra_nb6_QMC}. Since there is no direct Urca threshold for QMC-RMF3 EoS within the density range considered, we find that, similar to the results for IUF EoS at density below the direct Urca threshold, the bulk viscosity increases by five orders of magnitude in the suprathermal regime at $T=0.5$ MeV at both the densities $n_B=n_0,\,6n_0$. At temperature $T=1$ MeV the bulk viscosity increases by three orders of magnitude with the amplitude of oscillations.

Our results demonstrate that at low temperatures relevant in the relatively cooler cores of BNS mergers, bulk viscous dissipation can be enhanced by several orders of magnitude to become an effective damping mechanism for large-amplitude density oscillations. 
\section{Discussions}
\label{sec:discussions}
In this work, we investigated the behaviour of amplitude-dependent bulk viscous dissipation of density oscillations by consistently incorporating, for the first time, the thermal blurring effects and collisional broadening of in-medium nucleons into the equilibration channels using the NWA framework. We used two finite temperature EoS, IUF with a direct Urca threshold density at $n_B=4.1n_0$, and QMC-RMF3 which has no direct Urca threshold density. 

We found that the bulk viscous damping transitions into the suprathermal regime when the density oscillation amplitude reaches about ten percent of the background density. At low temperature, $T\lesssim 1$ MeV, this non-linear enhancement can increase the bulk viscosity by three to five orders of magnitude. This can provide a highly efficient damping mechanism for large-amplitude density oscillations particularly during the inspiral and early post-merger phases, or during the late-time cooling phase of the merger remnant or core-collapse supernova.
This can have direct implications on the composition and mass of ejecta from the BNS merger. Furthermore, to accurately predict the gravitational wave spectrum from BNS merger simulations it becomes important to consistently incorporate the effects of amplitude dependent bulk viscous damping of density oscillations. 

\section{acknowledgments}
The authors would like to acknowledge the usage of the High Performance Computing facility PEGASUS at Inter-University Centre for Astronomy and Astrophysics (IUCAA) for
the numerical calculations. The authors also thank the members of the NSGW group at IUCAA as well as Suprovo Ghosh for useful discussions.

\appendix 

\section{Appendix: Polynomial fit for net reaction rate}
\label{sec:appendix_a}
\begin{figure}[h]
    \centering
    \includegraphics[width=\linewidth]{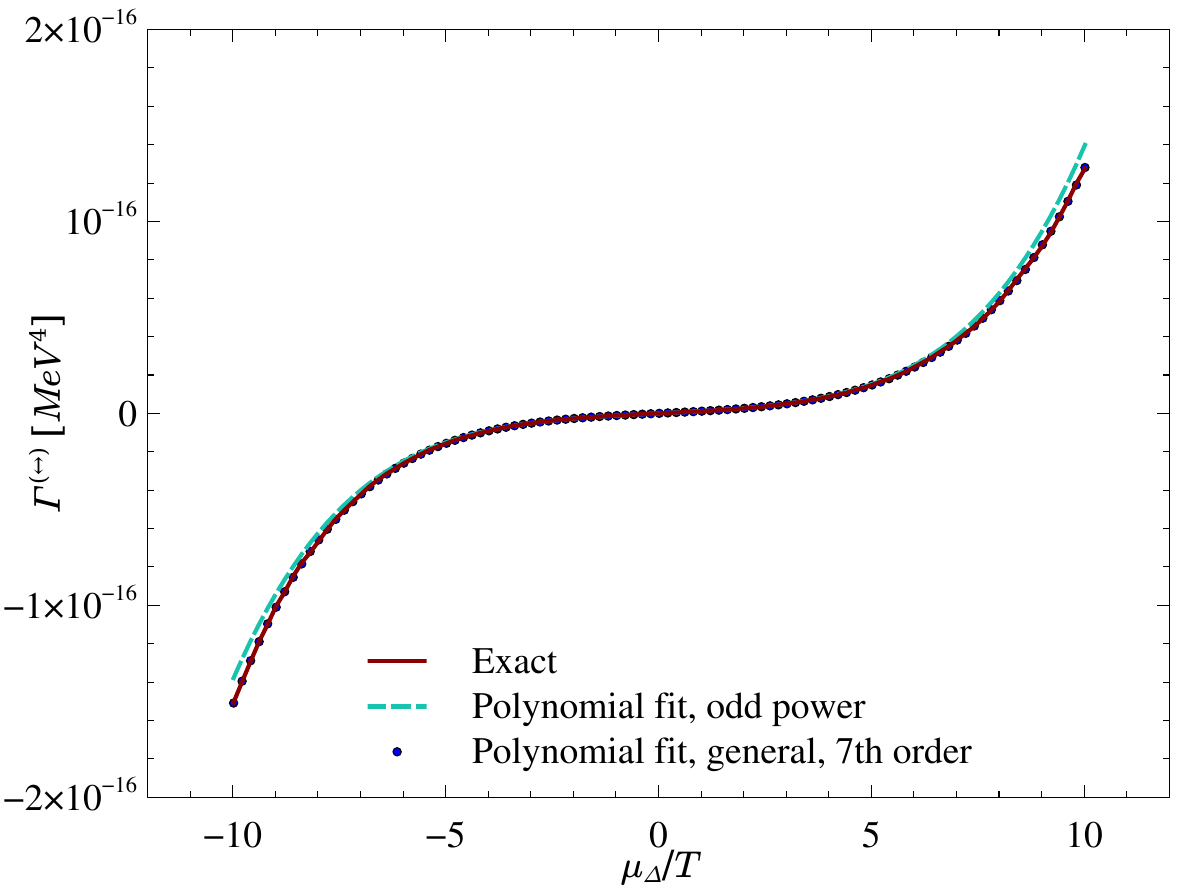}
    \caption{The net reaction rate as a function of deviation from true flavor equilibrium, $\mu_{\Delta}=\mu_n - \mu_p - \mu_e - \mu_I$ at $n_B=n_0$.}
    \label{fig:polyfit_rates}
\end{figure}

In Fig.~\ref{fig:polyfit_rates} in this section, we display a general polynomial fit for the net reaction rate as a function of $\mu_{\Delta}/T$ from Eq.~\ref{eq:rates_expansion} for IUF EoS at $T=1$ MeV. The polynomial fit with only odd powers in $\mu_{\Delta}/T$ (indicated by the black dotted line) following \cite{Alford:2010gw} does not accurately capture the thermal blurring and collisional broadening effects in the net reaction rates, so we employ a polynomial fit in all powers of $\mu_{\Delta}/T$ as demonstrated by the cyan line.

\bibliography{bibliography}{}



\end{document}